\documentclass[11pt]{article}

\usepackage[utf8]{inputenc}
\usepackage[english]{babel}
\usepackage[margin=2.3cm]{geometry} 
\usepackage{graphicx} 
\usepackage{amsthm, amsmath, amssymb} 
\usepackage{tcolorbox} 
\usepackage{placeins} 
\usepackage[numbers]{natbib} 
\usepackage{lineno}	
\usepackage{authblk} 
\usepackage{appendix} 

 \usepackage{setspace} 

\newcommand{\papertitle}{Structural build-up at rest in the induction and acceleration periods of OPC}

\newcommand{\storagemodulus}{G'}
\newcommand{\storagemodulusref}{\storagemodulus_0}
\newcommand{\storagemodulusnorm}{G^*}

\newcommand{\cumulheat}{H}

\newcommand{\cumulheatzero}{H_{0}}

\newcommand{\deltacumulheatHzero}{\cumulheat-\cumulheatzero}
\newcommand{\refhydrationamount}{H_0}

\newcommand{\interparticleinteraction}{\sigma}
\newcommand{\numcontacts}{N}

\newcommand{\solidvolfrac}{\phi}

\newcommand{\ie}{{\textit{i.e.}}}
\graphicspath{ {./} }
\begin{document}

\title{\papertitle}
\author{Luca Michel}
\author{Lex Reiter}
\author{Antoine Sanner}
\author{Robert J. Flatt}

\author{David S. Kammer}
\affil{Institute for Building Materials, ETH Zurich, Switzerland}

\maketitle

\section*{Abstract}
Structural build-up in fresh cement paste at rest is characterized by time evolutions of storage modulus and yield stress, which both increase linearly in time during the induction period of hydration, followed by an exponential evolution after entering the acceleration period.
While it is understood that C-S-H formation at contact points between cement particles dictates build-up in the acceleration period, the mechanism in the induction period lacks consensus.
Here, we provide experimental evidence that, at least in absence of admixtures, structural build-up at rest originates in both periods from the same mechanism.
We couple calorimetry and oscillatory shear measurements of OPC at different w/c ratios, capturing how the storage modulus evolves with changes in cumulative heat.
We obtain an exponential relation between stiffness and heat, with the same exponent in both the induction and acceleration periods.
This suggests that C-S-H formation dictates build-up at rest in both periods.


\section{Introduction}
When cement particles are suspended in water, they rapidly form an arrested network due to attractive interparticle interactions~\cite{flattDispersionForcesCement2004}.
The cement particles also undergo hydration reactions, forming products that strengthen the arrested network.
This combination of physical attractive interactions and chemical formation of hydration products results in an increase of macroscopic mechanical properties over time, a phenomenon referred to as structural build-up, or structuration~\cite{rousselOriginsThixotropyFresh2012, reiterStructuralBuildupDigital2019a}.
Structural build-up gained increasing interest in the past decade with the advent of self-compacting concrete and digital fabrication with concrete~\cite{rousselThixotropyModelFresh2006, wanglerDigitalConcreteOpportunities2016a}.
For example, in the case of layered extrusion, properly controlling the evolution of yield stress in each printed layer is key for successful printing~\cite{reiterRoleEarlyAge2018a, rousselRheologicalRequirementsPrintable2018}.
From a practical point of view, this type of challenge has been overcome in a number of projects~\cite{mechtcherineExtrusionbasedAdditiveManufacturing2020a, mennaOpportunitiesChallengesStructural2020, bosRealitiesAdditivelyManufactured2022}.
However, from a theoretical viewpoint, the mechanisms underlying structural build-up at rest are not fully understood, including for Ordinary Portland Cement (OPC).

\begin{figure}[ht!]
   \centering
   \includegraphics[width=0.7\linewidth]{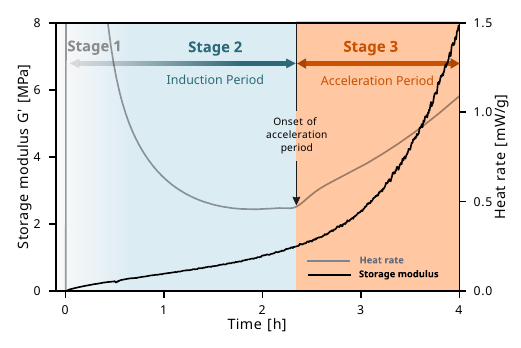} 
 \caption{\textit{Structural build-up at rest can be separated in three stages. Stage 1: initial non-linear evolution of the storage modulus. Stage 2: linear storage modulus time evolution in the induction period of cement hydration. Stage 3: exponential time evolution of storage modulus in the acceleration period of cement hydration. Data from an OPC 52.5R at w/c 0.46.}}
 \label{fig:p01:intro_3stages}
 \end{figure}

From a phenomenological perspective, early-age structural build-up in OPC can be separated into three stages~\cite{jiaoThixotropicStructuralBuildup2021, huangEvolutionElasticBehavior2020, zhangClarifyingQuantifyingDriving2023}. This separation is based on different characteristic time evolutions of storage modulus. By comparing the storage modulus evolution to heat rate on the same sample, it appears that the three stages in structural build-up overlap with the main hydration periods (see Figure~\ref{fig:p01:intro_3stages}).
While an analogous separation in three stages is obtained from yield stress measurements~\cite{lecompteNonlinearModelingYield2017}, we focus in this work on continuous measurements of storage modulus $\storagemodulus$.
The first stage is identified by a non-linear $\storagemodulus$ time evolution, which overlaps with the initial peak in heat rate data. 
In the second stage, the time evolution of $\storagemodulus$ is linear in time, over the induction period of cement hydration. 
The third stage is identified by an exponential evolution of $\storagemodulus$ over time, starting at the onset of the acceleration period.

 \FloatBarrier

Linear evolutions of storage modulus and yield stress in the induction period (stage 2) followed by exponential evolutions in the acceleration period (stage 3) have been reported in multiple studies~\cite{rousselOriginsThixotropyFresh2012, perrotPredictionLateralForm2015, lecompteNonlinearModelingYield2017, maExperimentalModelingStudy2018a}.
However, the mechanisms dictating build-up at rest are only understood for the third stage. Opinions still diverge regarding the underlying mechanisms in stage 2.

The third stage, which coincides with the acceleration period of cement hydration, is much better understood.
Because of the small size and low amounts of products formed, the number of contact points between cement grains remains unchanged \cite{flattDispersionForcesCement2004, rousselOriginsThixotropyFresh2012}.
Furthermore, ion-ion correlation forces between C-S-H particles result in strong attractive interactions~\cite{plassardNanoscaleExperimentalInvestigation2005}.
The cement particles being fixed in a dense flocculated network, the formation of C-S-H on their surface strengthens contacts between particles, dictating structural build-up at rest.
This was pictured by~\citet{rousselOriginsThixotropyFresh2012} as an increase in the number and size of C-S-H bridges resulting in a rigidification of the network, as also stated in a number of other studies~\cite{jiaoThixotropicStructuralBuildup2021, perrotPredictionLateralForm2015, lowkeThixotropySCCModel2018, plassardNanoscaleExperimentalInvestigation2005, jiangStudiesMechanismPhysicochemical1995, nonatInteractionsChemicalEvolution1994}.

For the second stage, various mechanisms have been proposed to explain structural build-up at rest.
Different studies~\cite{rousselOriginsThixotropyFresh2012, perrotPredictionLateralForm2015, lowkeThixotropySCCModel2018} suggest that C-S-H nuclei forming at contact points between cement particles dictate structural build-up in the induction period, implying that the same mechanism dictates build-up at rest in both the induction and acceleration periods.
In contrast, the evolution of storage modulus and yield stress in the same period was recently proposed to result from purely colloidal interactions~\cite{huangEvolutionElasticBehavior2020, zhangClarifyingQuantifyingDriving2023}.
Yet another hypothesis suggests that ettringite formation rather than C-S-H dictates build-up in the first hours of cement hydration~\cite{jakobRelatingEttringiteFormation2019}.
Thus, there is no strong consensus about the mechanism dictating structural build-up in the induction period.

Here, we provide evidence that the same mechanism dictates structural build-up at rest in both the induction and acceleration periods of OPC.
From coupled calorimetry and rheometry data, we demonstrate that the storage modulus varies as a single exponential function of the cumulative heat through both stages 2 and 3.
This relation suggests that the hydration component of structural build-up at rest in OPC mainly comes from the formation of C-S-H, as it is the main contributor to build-up in the acceleration period \cite{rousselOriginsThixotropyFresh2012, jiaoThixotropicStructuralBuildup2021, perrotPredictionLateralForm2015, plassardNanoscaleExperimentalInvestigation2005, jiangStudiesMechanismPhysicochemical1995}.
Early-age structural build-up in OPC can thus be described with two regimes, independent of hydration kinetics, rather than three as considered until now. Trying to directly explain structural build-up with respect to time is misleading because of the non-linear changes in hydration rate. It suggests that such approaches are almost certainly bound to fail because they cannot properly deconvolute hydration kinetics from the mechanisms responsible for structural build-up. This appears to largely explain the numerous discrepancies found in the literature.

\section{Coupling storage modulus and heat}
\begin{figure}[ht!]
  \centering
  \includegraphics[width=\linewidth]{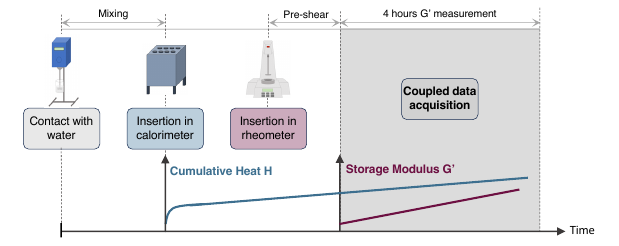}
\caption{\textit{Experimental workflow. For a given w/c ratio, we mix a paste with an IKA mixer for $3~\mathrm{minutes}$ at $500~\mathrm{RPM}$ following a procedure introduced in \cite{mantellatoRelatingEarlyHydration2019}. After mixing, we insert around $5~\mathrm{g}$ of paste in the calorimeter and the heat measurement is started. Directly after, we insert around $10~\mathrm{g}$ of paste in the rheometer. After a pre-shear event in the rheometer, the storage modulus measurement is started for a total duration of $4~\mathrm{hours}$. We acquire calorimetry and rheometry data simultaneously over the entire storage modulus measurement period.}}
\label{fig:p01:experimental_workflow}
\end{figure}

Experimentally, structural build-up is usually characterised by time evolutions of yield stress or storage modulus~\cite{rousselOriginsThixotropyFresh2012, jiaoThixotropicStructuralBuildup2021}. 
For storage modulus measurements, the flocculated dense suspension is probed in its linear elastic regime, while for yield stress measurements the suspension's network has to be destroyed to bring the material to yield, pushing it beyond its linear elastic limit.
While yield stress and storage modulus characterize different regimes of deformation, they both evolve monotonically over time and have been reported to follow similar trends \cite{yuanMeasurementEvolutionStructural2017}.
In this work, we use storage modulus measurements to track structural build-up at rest.
This allows a non-destructive and, more importantly, continuous data acquisition.
In parallel to storage modulus measurements, we acquire isothermal calorimetry data.
For a given w/c investigated, we mix the paste. Part of it is then inserted in the calorimeter, and part of it is inserted in the rheometer, as summarized in Figure \ref{fig:p01:experimental_workflow}.
The \textit{simultaneous} recording of heat and storage modulus data on two instances of the same sample allows us to link chemical changes from hydration reactions to evolutions in macroscopic mechanical properties.
\FloatBarrier

The cement used is an OPC CEM I 52.5R (Holcim Normo 5R) cement. All the mixes are prepared with $30~\mathrm{g}$ of cement and distilled water. The water to cement ratios investigated are 0.36, 0.38, 0.40 and 0.42. The mixer used is an IKA Eurostar EURO-ST P CV mixer with a 4-bladed propeller stirrer. The dry powder is first added to a fraction of the total mixing water corresponding to a lower bound w/c ratio (in this case 0.34) and mixed for $1~\mathrm{minute}$ at $500~\mathrm{RPM}$. The remaining mixing water needed to reach the targeted w/c ratios is then added and the paste is mixed for 2~more minutes at 500 RPM. This is done in order to reduce self-mixing energy effects, as proposed in~\citet{mantellatoRelatingEarlyHydration2019}. The heat released by the hydration reactions is monitored with a TAM AIR isothermal calorimeter at 23C. Measurements are made with glass vials filled with about $5~\mathrm{g}$ of cement paste. Rotational rheometer measurements are performed on an Anton Paar Physica MCR 501 rheometer in parallel plate geometry using serrated plates on $10~\mathrm{g}$ of paste. A plastic hood limits sample drying over the measurement period. The measurement is strain controlled, except for a stress controlled step at the liquid-solid transition of the material following the pre-shear event. The gap between the upper and lower plate is $1~\mathrm{mm}$ and their diameters are $25~\mathrm{mm}$ and $50~\mathrm{mm}$ respectively. The rheometer measurement starts directly after the start of the calorimeter measurement.
The rheometer protocol is composed of a LAOS pre-shear of $30~\mathrm{seconds}$ at $10~\mathrm{\%}$ amplitude and $1~\mathrm{Hz}$ frequency, followed by a stress-controlled rest period where the shear stress is set to $\tau=0$ for $30~\mathrm{seconds}$. After these steps the sample is considered to be in a homogeneous state and the storage modulus measurement (SAOS) is started at an amplitude of $0.003~\mathrm{\%}$ and frequency of $1~\mathrm{Hz}$, and lasts for $4~\mathrm{hours}$.
Amplitude sweeps allowing to assess the linear elastic range of the pastes investigated are given in Appendix~\ref{appendix:p01:assess_ler}.
The storage modulus data is post-processed by applying a low-pass filter with cutoff~0.012.

\section{Results and Discussion}
\subsection{Experimental data}
\begin{figure}[ht!]
  \centering
  \includegraphics[width=0.9\linewidth]{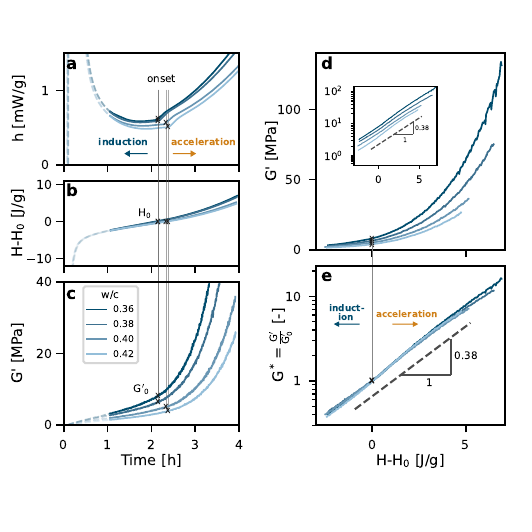}
\caption{\textit{Coupling calorimetry and rheometry data for OPC samples at different w/c ratios. \textbf{a)}~\textbf{Heat rate} over time from calorimetry measurements at different w/c ratios. The onset of the acceleration period is identified at the bump of heat rate data and marked by black cross. The first hour of data, plotted as dashed line, is discarded due to noise coming from sample insertion in the calorimeter. \textbf{b)}~\textbf{Change in cumulative heat} over time computed as the time integral of heat rate data starting at 1 hour. $\cumulheatzero$ is the change in cumulative heat at the onset of the acceleration period. \textbf{c)}~\textbf{Storage modulus} from small amplitude oscillatory shear measurements at different w/c ratios. $\storagemodulusref$ is the storage modulus at the onset of the acceleration period as marked by black cross. \textbf{d)}~\textbf{Coupling} storage modulus and cumulative heat data at different w/c ratios. Onset as identified from heat rate in (a) is marked as black cross. (inset) Semi-log plot for same data. \textbf{e)} \textbf{Normalized storage modulus} $\storagemodulus/\storagemodulusref$ plotted against change in cumulative heat $\deltacumulheatHzero$.}}
\label{fig:p01:master_curve}
\end{figure}

We conduct experiments on four different w/c ratios and record heat rate, cumulative heat, and storage modulus data over time.
The heat rate data allow us to identify the different periods of cement hydration, namely the initial peak, induction period, and acceleration period (see Fig.~\ref{fig:p01:master_curve}a).
More specifically, we identify the onset of the acceleration by the "bump" in heat rate seen in Figure~\ref{fig:p01:master_curve}a, which is probably linked to Portlandite nucleation.
The sample insertion in the calorimeter results in noise in the calorimetry data, impacting the initial peak of heat rate. We thus discard the first hour of calorimetry data (dashed lines in Figure~\ref{fig:p01:master_curve}a and \ref{fig:p01:master_curve}b).
Details on how we assess the time needed to dissipate this initial noise are given in appendix~\ref{appendix:p01:initial_calo_noise}.
The cumulative heat $\cumulheat$ captures the total heat generated by hydration reactions from a given point in time, and is the time integral of the heat rate.
We subtract the cumulative heat at the onset of the acceleration period $\cumulheatzero$ from $\cumulheat$, giving changes in cumulative heat $\deltacumulheatHzero$ (see Figure~\ref{fig:p01:master_curve}b).
Noteworthy, we normalize the heat rate as well as the cumulative heat data by mass of cement paste.

The storage modulus quantifies the evolution of stiffness of the network of cement particles, and increases monotonically over time for all the samples investigated, as shown in Figure~\ref{fig:p01:master_curve}c. We discard the first hour of storage modulus data (dashed lines in Figure~\ref{fig:p01:master_curve}c) as it cannot be coupled with unreliable calorimetry data.

The sorting of heat rate curves with w/c in Figures~\ref{fig:p01:master_curve}a and \ref{fig:p01:master_curve}b is the signature of different hydration kinetics taking place because of the filler effect~\cite{berodierUnderstandingFillerEffect2014}.
During mixing, the presence of neighbouring particles changes the shearing forces in the suspension.
This affects the dissolution, accelerating the kinetics in denser systems~\cite{juilland_dissolution_2010, juilland_effect_2012}.

Similarly to heat rate data, time evolutions of storage modulus are sorted with w/c ratio, as seen in Figure~\ref{fig:p01:master_curve}c. However, here, the sorting with w/c ratio does not solely result from different hydration kinetics.
For a given particle size distribution, denser systems will have more interparticle contacts per unit volume, resulting in higher values of $\storagemodulus$~\cite{bonacciContactMacroscopicAgeing2020}.
On top of this mechanical contribution, different hydration kinetics in samples at different w/c ratios can influence the evolution of $\storagemodulus$ in time.
Two pastes with the same chemical composition and same network of suspended particles, for a given \textit{hydration extent}, are expected to have the same storage modulus. However, if one of these pastes reacts faster than the other, at a given \textit{point in time}, it will have a higher stiffness.
It is thus unclear to what extent the sorting of the storage modulus curves in Figure~\ref{fig:p01:master_curve}c is impacted by different hydration kinetics across the samples.

We eliminate the effect of different hydration kinetics by directly considering the storage modulus evolution $\storagemodulus$ with respect to the change in cumulative heat $\deltacumulheatHzero$, as shown in Figure~\ref{fig:p01:master_curve}d.
By doing so, we can investigate how the storage modulus changes for a given extent of hydration, irrespective of the reaction kinetics.
We obtain a sorting of the $\storagemodulus$ vs.  $\deltacumulheatHzero$ curves with w/c ratio.
Furthermore, the $\storagemodulus$ vs. $\deltacumulheatHzero$ relations are exponential, as shown by the inset in Figure~\ref{fig:p01:master_curve}d.
The constant slope in the inset of Figure~\ref{fig:p01:master_curve} indicates that the same exponent holds for the different w/c investigated.
Thus, the following relation holds through both the induction and acceleration periods
\begin{equation}
\label{eq:p01:exponential_GDH}
    \storagemodulus = a e^{b (\deltacumulheatHzero)} ~,
\end{equation}
where $a$ is dependent on w/c ratio and $b$ has a value of 0.38 for all w/c ratios.

The dependence of $\storagemodulus$ on w/c ratio is canceled out when normalizing the storage modulus evolutions by the storage modulus value at the onset of the acceleration period, $\storagemodulusref$ (see Figures~\ref{fig:p01:master_curve}a and~\ref{fig:p01:master_curve}c). Normalizing the $\storagemodulus$ vs. $\deltacumulheatHzero$ evolutions by $\storagemodulusref$ causes all the curves at different w/c ratios to collapse on a single master curve, as shown in Figure~\ref{fig:p01:master_curve}e.
The resulting normalized storage modulus $\storagemodulusnorm$ evolutions are exponential with changes in cumulative heat $\deltacumulheatHzero$ through both the induction and acceleration periods of hydration, and can be described by the following relation
\begin{equation}
\label{eq:p01:exponential_GstarDH}
    \storagemodulusnorm = \frac{\storagemodulus}{\storagemodulusref} = e^{b(\deltacumulheatHzero)} ~,
\end{equation}
where $\storagemodulusref$ is the storage modulus at the onset of the acceleration period.
We provide data to assess the reproducibility of the approach in Appendix~\ref{appendix:p01:reproducibility_calorheo} and discuss the dependence of $\storagemodulusref$ on solid volume fraction in Appendix~\ref{appendix:p01:G0_phi}.

\FloatBarrier

\subsection{A simple mathematical model for contacts strengthening}
Generally, the strength of a dense suspension depends on interparticle forces, volume fraction, and size distribution of the particles.
Different studies have shown that these effects are independent of each other~\cite{zhouChemicalPhysicalControl2001, zhouYieldStressConcentrated1999, flattYodelYieldStress2006}.
For a given particle size distribution, varying the volume fraction of particles in the suspension results in changes to the initial number of contact points between particles $N$.
In this work, all measurements are performed on the same OPC with a fixed particle size distribution.
Hence, we express the storage modulus of a cement paste as the product of a function of the total number of contact points between cement grains $f(N)$ and the stiffness of these grain-to-grain contacts $\interparticleinteraction$
\begin{equation}
G' \propto f(\numcontacts) \cdot \interparticleinteraction ~.
\end{equation}

Furthermore, we assume that hydration products only stiffen contacts, leaving $\numcontacts$ unchanged.
While it is unclear whether very early hydration products impact the initial number of contacts $\numcontacts$, the products formed over the time periods considered in this work are much smaller than the characteristic size of cement grains~\cite{flattDispersionForcesCement2004}.
The role of the percolation threshold on yield stress was recently highlighted in cement pastes containing chemical admixtures~\cite{sha_superplasticizers_2023}. Due to the absence of admixtures in the samples investigated here, and given that no remixing occurs over the $\storagemodulus$ measurement, we assume that the ongoing hydration does not impact the percolation threshold, leaving $f(\numcontacts)$ unchanged.
Following the point of view of~\citet{rousselOriginsThixotropyFresh2012}, we thus assume that, over the time period considered in this study, hydration reactions stiffen the network of cement grains without changing the number of contact points between the grains.
Under these assumptions, the evolution of storage modulus with hydration becomes
\begin{equation}
\label{eq:p01:GfNsigmaH}
\storagemodulus(\cumulheat) \propto f(\numcontacts)\cdot \interparticleinteraction(\cumulheat) ~,
\end{equation}
where $\interparticleinteraction(\cumulheat)$ describes the stiffness of contacts between cement grains at a given hydration amount $\cumulheat$.

The geometrical term $f(\numcontacts)$ in equation \ref{eq:p01:GfNsigmaH} is canceled out when normalizing $\storagemodulus(\cumulheat)$ by a reference storage modulus $\storagemodulusref$, which is defined as the storage modulus at the point where hydration has advanced up to an amount $\refhydrationamount$
\begin{equation}
\storagemodulusref = \storagemodulus(\refhydrationamount) \propto f(\numcontacts) \cdot \interparticleinteraction(\refhydrationamount) ~.
\end{equation}
The normalized storage modulus $\storagemodulusnorm$ is thus given by
\begin{equation}
\label{eq:p01:Gstar}
\storagemodulusnorm (\cumulheat) =  \frac{\storagemodulus(\cumulheat)}{\storagemodulusref} = \frac{\interparticleinteraction(\cumulheat)} {\interparticleinteraction(\refhydrationamount)} ~,
\end{equation}
and is indeed independent of the number $\numcontacts$ of contacts between cement grains.

\subsection{Comparing the model with experimental data}
We first consider the relation between storage modulus and heat of hydration.
In the model, we assume that hydration only impacts existing grain-to-grain contact points, without changing their total number $\numcontacts$.  
This results in a separation of $\storagemodulus$ in a geometrical term $f(\numcontacts)$ and a stiffening term $\interparticleinteraction(\cumulheat)$ in equation~\ref{eq:p01:GfNsigmaH}.
This functional form is the same as obtained from the experimental data (see Figure~\ref{fig:p01:master_curve}d and equation~\ref{eq:p01:exponential_GDH}).
We thus link the parameter $a$ preceding the exponential term in equation~\ref{eq:p01:exponential_GDH} to the geometrical term $f(\numcontacts)$.
$a$ captures the network of suspended cement grains. It is dependent on w/c ratio as predicted by the model and observed experimentally.
Similarly, we link the exponential term in equation~\ref{eq:p01:exponential_GDH} to the stiffening term $\interparticleinteraction(\cumulheat)$.
As $\interparticleinteraction(\cumulheat)$ describes the stiffening of single grain-to-grain contacts, it is independent of w/c ratio. This corresponds to our experimental results, where all the $\storagemodulus$ vs. $\deltacumulheatHzero$ curves in Figure~\ref{fig:p01:master_curve}d have the same exponent.
The parameter $b$ in equation~\ref{eq:p01:exponential_GDH} thus captures the stiffening mechanism of single contacts between cement grains.

According to the model, the normalized storage modulus $\storagemodulusnorm$ is independent of the geometry of the network of suspended particles and describes the stiffening mechanism of single contacts.
It relies on the storage modulus at a reference hydration state.
Identifying a chemical reference state from time evolutions of heat is challenging because of different hydration kinetics across samples at different w/c ratio.
In this regard, we take the reference storage modulus $\storagemodulusref$ at the onset of the acceleration period. The onset marks a transition from a slow dissolution and precipitation of products in the induction period, to a much faster and accelerating dissolution and precipitation in the acceleration period \cite{scrivenerAdvancesUnderstandingCement2019}.
Even if the onset takes place at different points in time, at the onset, samples at different w/c ratios will all be undergoing the same chemical transition, which we believe can serve as reference hydration state.
Computing the normalized storage modulus $\storagemodulusnorm$ experimentally results in a single exponential evolution with changes in heat (see Figure~\ref{fig:p01:master_curve}e and equation~\ref{eq:p01:exponential_GstarDH}). This result is in line with the prediction of the model. The normalized storage modulus $\storagemodulusnorm$ thus appears to capture the stiffening mechanism of single contacts between cement grains compared to a reference state, and can be used to describe structural build-up at rest.
We note here that, because of the exponential dependence of storage modulus on heat, the reference storage modulus $\storagemodulusref$ does not necessarily need to be taken at the onset of the acceleration period.
A collapse of the $\storagemodulus$ vs. $\deltacumulheatHzero$ curves is obtained for any value of $\storagemodulus$ corresponding to a fixed hydration state through the samples.
We stress here that this freedom in choice of reference only holds for storage modulus data expressed in terms of heat, not time.

The good agreement of our experimental data with the presented model suggests that structural build-up at rest, over the time periods investigated here, results from a strengthening of contact points between cement grains, leaving the total number of grain-to-grain contacts unchanged. 

\subsection{Mechanisms dictating build-up at rest}

Structural build-up at rest can be described by evolutions of the normalized storage modulus $\storagemodulusnorm$ with changes in cumulative heat $\deltacumulheatHzero$.
As discussed above, $\storagemodulusnorm$ evolutions likely capture the strengthening mechanism of contact points between cement particles.
From Figure~\ref{fig:p01:master_curve}d, it appears that the same exponential dependence of $\storagemodulusnorm$ on heat holds through both the induction and acceleration periods of cement hydration.
These results suggest that the same mechanism dictates build-up at rest in both periods of hydration.
Since C-S-H is the main contributor to structural build-up in the acceleration period~\cite{rousselOriginsThixotropyFresh2012}, we infer that it also plays a dominant role in the induction period.
After the heat of the first hydration peak has dissipated, structural build-up at rest in OPC is dictated by formation of C-S-H strengthening contact points between cement particles in the paste.
It should be noted that our results only cover the initial part of the acceleration period, limited by the range of stiffness that can be probed in a rheometer.

While the $\storagemodulusnorm$ master curve hints that C-S-H is the main contributor to structural build-up in the induction and acceleration periods, it does not allow to directly draw conclusions with regard to other hydration products such as Portlandite and ettringite, or changes in interparticle forces.
The present $\storagemodulusnorm$ results are nevertheless in line with statements by 
\citet{rousselOriginsThixotropyFresh2012}, who proposed that in the induction period, the network of cement particles is strengthened by C-S-H nuclei forming at contact points between cement grains.
In the acceleration period, they pictured the strengthening as resulting in an increase in the number and size of C-S-H bridges at the contact points between cement grains.
While \citet{rousselOriginsThixotropyFresh2012} make a distinction in both periods based on the kinetics of formation, our results suggest that the underlying physics of contact strengthening is the same, no matter how fast C-S-H is formed. We also infer that fitting strengthening data as a function of time can be misleading because the non-linear hydration kinetics can easily be misinterpreted as a change of contact strengthening mechanism, when there is none, but just a change in hydration kinetics.

The present results contrast findings by \citet{jakobRelatingEttringiteFormation2019}, who suggested that ettringite formation dictates structural build-up at rest in the induction period of OPC.
Our results show that storage modulus evolutions with heat are dictated by the same process in both the induction and acceleration periods of hydration.
If different hydration products were responsible for the storage modulus evolution in the induction and acceleration period, we would expect two different functional forms before and after the onset in Figure~\ref{fig:p01:master_curve}e.
Nevertheless, ettringite likely does play a role in structural build-up at rest, but not in the time ranges investigated here. 
We infer that ettringite only impacts structural build-up in the first minutes of hydration, a time period not accessible with the present experimental setup.
Similarly, the flocculation processes taking place upon contact of cement particles with water are not observed in the $\storagemodulusnorm$ vs. $\deltacumulheatHzero$ curves. These are already completed by the time we start the coupling of calorimetry and rheometry data.

In-situ isothermal calorimetry allows to record heat data without the important noise arising from sample insertion in the calorimeter. However, we do not recommend this approach for coupled calorimetry and rheometry measurements, as the paste in the calorimeter would be subjected to a different mixing than the paste in the rheometer, strongly affecting the hydration reactions.
In that case, the heat data recorded in the calorimeter would not be representative of the hydration reactions taking place in the rheometer.

The proposed normalized storage modulus $\storagemodulusnorm$ is an alternative modeling approach to existing phenomenological models describing structural build-up, such as the linear model by \citet{rousselThixotropyModelFresh2006} or the exponential model by \citet{perrotPredictionLateralForm2015}.
These models are based on time evolutions of yield stress and are thus handy for practical applications. 
Nevertheless, the normalized storage modulus $\storagemodulusnorm$ depends on heat evolutions, directly capturing the effect of hydration on build-up.
Furthermore, we understand $\storagemodulusnorm$ as a description of the stiffening of granular contact points in the suspension, compared to a reference state.
We believe that the normalized storage modulus $\storagemodulusnorm$ can be used in addition to existing phenomenological models to gain further fundamental understanding, especially in more complex systems such as blended cements.

\begin{figure}[ht!]
  \centering
  \includegraphics[width=1\linewidth]{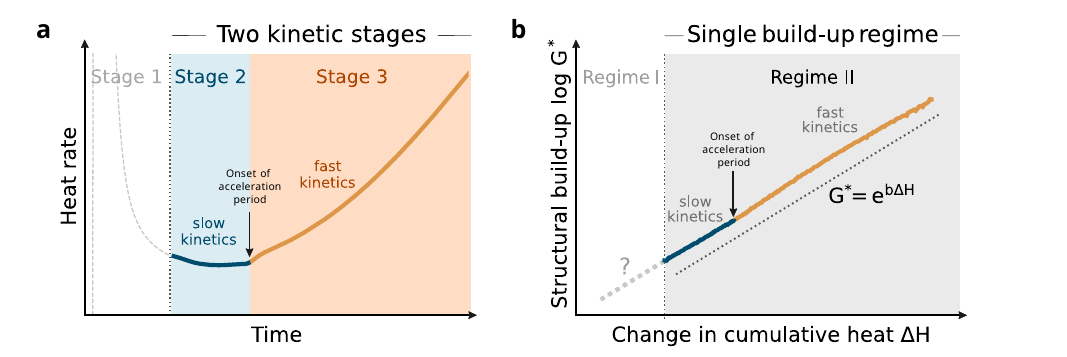}
\caption{\textit{Two distinct hydration kinetics stages (stages 2 and 3) correspond to a single structural build-up regime (regime II). \textbf{a)}~\textbf{Heat rate} data over time. Stage 1 is the initial peak, capturing initial dissolution and noise from sample insertion in the calorimeter. The second stage corresponds to the induction period of hydration, whereby a slow hydration kinetics is observed. Stage 3 is the acceleration period of cement hydration, where a faster hydration kinetics occurs. \textbf{b)}~\textbf{Structural build-up} evolution with changes in cumulative heat. Structural build-up is described with the normalized storage modulus $\storagemodulusnorm$. Despite the different hydration kinetics observed in stages 2 and 3, a single regime of $\storagemodulusnorm$ evolution with changes in cumulative heat is observed (regime II). The first structural build-up regime (regime I) cannot be accessed with the current setup due to noise in the kinetic stage 1.}}
\label{fig:p01:fig4}
\end{figure}

From a more general perspective, the separation of structural build-up in three stages presented in Figure \ref{fig:p01:intro_3stages} can be simplified to two main regimes, whereby the kinetic stages 2 and 3 can be merged in a single build-up regime (see Figures~\ref{fig:p01:fig4}a and \ref{fig:p01:fig4}b).
The key finding of this paper is therefore that although the hydration kinetics changes between stage 2 and 3, the fundamental mechanism responsible for structural build-up, probably the formation of C-S-H, does not change. It just takes place at a different rate because the chemistry is happening at a different rate. 

The mechanisms taking place prior to the induction period (regime 1 in Figure~\ref{fig:p01:fig4}b) cannot be elucidated with our approach. The reactions occurring in that period are complex and still under debate. Furthermore, the heat released includes in large part the initial reaction of C$_\mathrm{3}$A, and is difficult to measure experimentally. However, C-S-H \ appears to dictate structural build-up after this first regime, independent of the reaction kinetics (see Figure~\ref{fig:p01:fig4}b).
It thus appears essential to examine the extent to which this also occurs in stage 1 in future studies.


\FloatBarrier

In systems including chemical admixtures, the situation could be more complex. In such cases, any additional surface will potentially contribute to reducing the surface coverage by chemical admixtures \cite{sha_superplasticizers_2023, mantellato_shifting_2020}. This a priori also modifies the network stiffness, which however remains influenced by direct contract strengthening by hydrates forming between grains \cite{mantellato_flow_2017}. The present study, however, lays a good basis to address these more complex questions, underlying how easy it can be to misinterpret data if strengthening is simply fit as a function of time and separated in semi-arbitrarily defined periods.

\section{Conclusion}
In this work, we investigated structural build-up at rest in OPC by considering the storage modulus evolutions with respect to changes in cumulative heat at different w/c ratios and in absence of chemical admixtures.
This approach led us to the following conclusions.

First, structural build-up appears to be dictated by the same mechanism in both the induction and acceleration periods of hydration. 
Storage modulus evolutions with changes in cumulative heat result in a single exponential evolution throughout both periods. 
This contrasts with the two distinct stages - linear in the induction period and exponential in the acceleration period - observed from time-resolved storage modulus data.
It suggests that structural build-up at rest can be described with two regimes, rather than three stages as considered until now.
The approach presented here shows that describing structural build-up from time evolutions can be misleading because of the non-linear changes in hydration rate. Investigating structural build-up from heat evolutions allows us to deconvolute hydration kinetics from the mechanisms underlying structural build-up.
However, the first stage of reaction requires further investigation, because the current approach cannot deconvolute heat combining initial dissolution of C$_\mathrm{3}$A and C$_\mathrm{3}$S, in addition to transition effects linked to inserting the sample in the calorimeter.

Furthermore, structural build-up at rest appears to be dictated by a strengthening of contact points by C-S-H, as this is the main product forming in the acceleration period.
By coupling storage modulus and changes in cumulative heat, different hydration kinetics in samples at different w/c ratios are canceled out.
This allows us to compare build-up in samples at different w/c ratios.
We observe that varying the w/c ratio impacts the number of contact points, without changing the strengthening mechanism.

Finally, we introduced the normalized storage modulus $\storagemodulusnorm$, capturing the stiffening of contact points between cement grains in the paste.
We believe that $\storagemodulusnorm$ can be further used to investigate more complex systems, such as blended cements.

\section{CRediT authorship contribution statement}
\textbf{Luca Michel}: Methodology, Investigation, Formal Analysis, Data Curation and Visualisation, Writing Original Draft.\\
\textbf{Lex Reiter}: Conceptualisation, Supervision, Methodology, Formal Analysis, Writing-Review and Editing\\
\textbf{Antoine Sanner}: Formal Analysis, Writing- Review and Editing\\
\textbf{Robert J. Flatt}: Conceptualisation, Supervision, Resources, Formal Analysis, Writing- Review and Editing.\\
\textbf{David S. Kammer}: Conceptualisation, Supervision, Formal Analysis, Writing- Review and Editing, Project administration, Funding acquisition.

\section{Declaration of competing interest}
The authors declare that they have no known competing financial interests or personal relationships that could have appeared to influence the work reported in this paper.

\section{Acknowledgements}
The authors acknowledge Franco Zunino for useful discussions, as well as Arnesh Das for help during the method development.
The authors also acknowledge Michal Hlobil for feedback on the manuscript draft.
The authors acknowledge the Swiss National Science Foundation for financial support under grant number 200021\_200343.

\appendix

\section{Assessing the linear elastic range for rheometer measurements}
\label{appendix:p01:assess_ler}
\begin{figure}[ht!]
  \centering
  \includegraphics[width=0.7\linewidth]{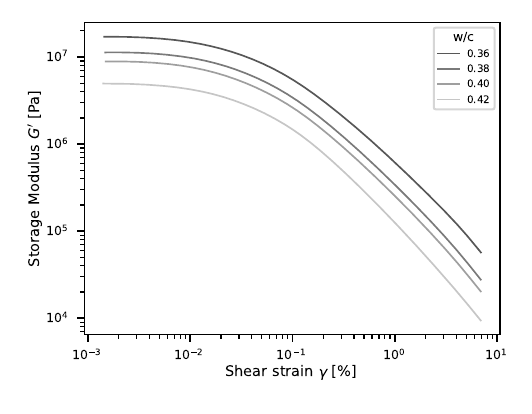}
\caption{\textit{Amplitude sweeps on OPC samples at different w/c ratios. Frequency 1 Hz.}}
\label{fig:p01:assess_ler}
\end{figure}

Results from amplitude sweeps are reported in Figure~\ref{fig:p01:assess_ler}.
Based on these results, the amplitude of the storage modulus measurements at rest is taken as 0.003\% in order to probe the materials in their linear elastic range.

\FloatBarrier

\section{Initial noise in calorimetry data}
\label{appendix:p01:initial_calo_noise}
\begin{figure}[ht!]
  \centering
  \includegraphics[width=0.7\linewidth]{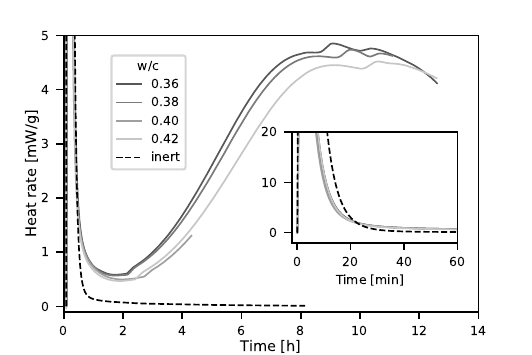}
\caption{\textit{A comparison of isothermal calorimetry data from OPC at different w/c ratios with an inert sample (dashed line) allows to assess how much time is needed to dissipate the noise in the heat signal coming from the sample insertion in the calorimeter. }}
\label{fig:p01:noise_calo}
\end{figure}

The noise in the first peak of heat rate caused by sample insertion can be assessed by measuring an inert sample, as shown in Figure \ref{fig:p01:noise_calo}.
The inert sample is a blend of 70\% metakaolin and 30\% limestone in 0.2M NaOH at w/c 0.83. It is inserted in a TamAir isothermal calorimeter at 23C in the same way as the OPC samples. After about 40 minutes, the noise in the heat signal coming from the sample insertion has decayed to 0 and no heat is generated by the inert sample.
In this work, we consider heat data from one hour onwards, ensuring that the data is not biased by the initial noise coming from the sample insertion in the calorimeter.

\FloatBarrier

\section{Reproducibility of coupled calorimetry-rheometry measurements}
\label{appendix:p01:reproducibility_calorheo}
\begin{figure}[ht!]
  \centering
  \includegraphics[width=0.7\linewidth]{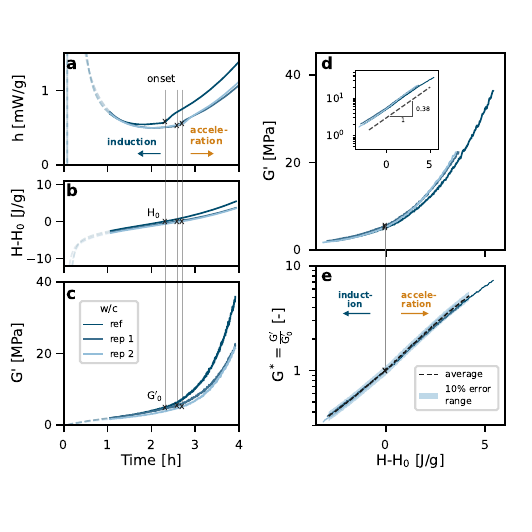}
\caption{\textit{Assessing the reproducibility of the measurements on an OPC w/c 0.4 sample. a) Time evolutions of heat rate. b) Time evolutions of storage modulus. c) Coupled storage modulus and heat data. d) Normalized storage modulus vs. heat data. After coupling, all the curves collapse within 10\% of error.}}
\label{fig:p01:reproducibility}
\end{figure}

The reproducibility of the coupled calorimetry and rheometry measurements is shown in Figure \ref{fig:p01:reproducibility}, where 3 measurements on an OPC at w/c 0.4 are shown.
The time evolutions of heat rate and storage modulus data show poor reproducibility in Figures \ref{fig:p01:reproducibility}a and \ref{fig:p01:reproducibility}b.
However, when normalized storage modulus and heat data are coupled, all repetitions fall within an error range of 10\%. 

\FloatBarrier

\section{Dependence of $\storagemodulusref$ on solid volume fraction}
\label{appendix:p01:G0_phi}
\begin{figure}[ht!]
  \centering
  \includegraphics[width=0.7\linewidth]{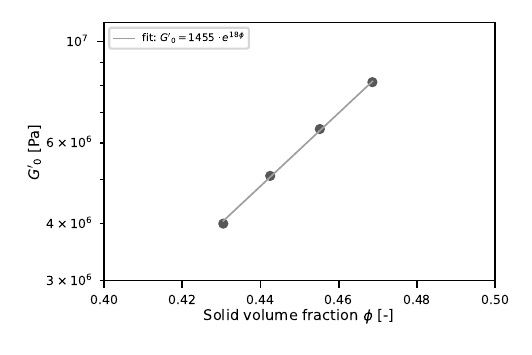}
\caption{\textit{Dependence of the reference storage modulus $\storagemodulusref$ on solid volume fraction. Note the logarithmic scale on the y axis.}}
\label{fig:p01:G0_phi}
\end{figure}

The dependence of the reference storage modulus at the onset of the acceleration period $\storagemodulusref$ on initial solid volume fraction $\phi$ is given in Figure~\ref{fig:p01:G0_phi}. Low values of $\solidvolfrac$ correspond to high w/c ratios (dilute samples). The relation is expected to diverge at the maximum packing fraction. The narrow range of solid volume fractions makes it difficult to assess the quality of different fits.
We nevertheless fit the data with an exponential relation, as already done in~\cite{reiterRoleEarlyAge2018a}.
Exponential dependence of yield stress on solid volume fraction was reported in~\citet{mantellato_shifting_2020}.
Note that we consider the initial solid volume fraction, \ie, the volume fraction of unhydrated cement grains. We do so because changes in volume fraction over the time periods investigated in this study are either not accessible experimentally (very early hydration products)  or negligible (C-S-H and Portlandite formed in the induction period).

\FloatBarrier

\bibliography{00_p01.bib}

\end{document}